\documentclass[utf8]{frontiersSCNS} 

\usepackage{url,hyperref,lineno,microtype,subcaption}
\usepackage[onehalfspacing]{setspace}
\usepackage{multicol}
\usepackage{color}

\def\keyFont{\fontsize{8}{11}\helveticabold }
\def\firstAuthorLast {Wo\l owska {et~al.}} 
\def\Authors{ Aleksandra Wo\l owska\,$^{1,*}$, Magdalena Kunert-Bajraszewska\,$^{2}$, Kunal Mooley\,$^{2}$ and Gregg Hallinan\,$^{3}$}

\def\Address{$^{1}$Toru\'n Centre for Astronomy, Faculty of Physics, Astronomy and Informatics, NCU, Grudziadzka 5 , 87-100 Toru\'n, Poland \\
$^{2}$Centre for Astrophysical Surveys, University of Oxford, Denys Wilkinson Building, Keble Road, Oxford OX1 3RH, UK \\
$^{3}$ Cahill Center for Astronomy, MC 249-17, California Institute of Technology, Pasadena, CA 91125, USA}
\def\corrAuthor{Aleksandra Wo\l owska}

\def\corrEmail{}

\begin{document}
\onecolumn
\firstpage{1}

\title[Changing-look AGNs or short-lived radio sources?]{Changing-look AGNs or short-lived radio sources?} 

\author[\firstAuthorLast ]{\Authors} 
\address{} 
\correspondance{} 

\extraAuth{}

\maketitle

\begin{abstract}

\section{}
The evolution of extragalactic radio sources has been a fundamental problem in the study of active galactic nuclei for many years. A standard evolutionary model has been created based on observations of a wide range of radio sources.
In the general scenario of the evolution, the younger and smaller Gigahertz-Peaked Spectrum (GPS) and Compact Steep Spectrum (CSS) sources become large-scale FRI and FRII objects. However, a growing number of observations of low power radio sources suggests that the model cannot explain all their properties and there are still some aspects of the evolutionary path that remain unclear. There are indications, that some sources may be short-lived objects on timescales of $10^4$ - $10^5$ years. Those objects represent a new population of active galaxies. Here, we present the discovery of several radio transient sources on timescales of 5-20 years, largely associated with renewed AGN (Active Galactic Nucleus) activity. These changing-look AGNs possibly represent behaviour typical for many active galaxies.

\tiny
 \keyFont{ \section{Keywords:} galaxies, active-galaxies, evolution, quasars, recurrent-activity}
\end{abstract}

\section{Introduction}

The radio emission of extragalactic sources is usually explained by the existence of strong jets emitting non-thermal synchrotron radiation. However, AGNs with powerful jets represent only a small fraction of the entire population (Kellermann et al. 1989). The vast majority of sources are much less powerful in radio domain or even defined as radio-quiet. But even then they still emit radio waves at very small fluxes, compared to radio-loud AGNs. The relatively small amount of radio emission in some AGNs may be due the presence of less powerful jets (Ulvestad et al. 2005). This in turn connects the phenomena with the accretion process which can be radiatively inefficient in low radio luminosity AGNs (Merloni et al. 2003, Best \& Heckman 2012). 


\begin{figure}[t]
\center
\includegraphics[scale=0.5]{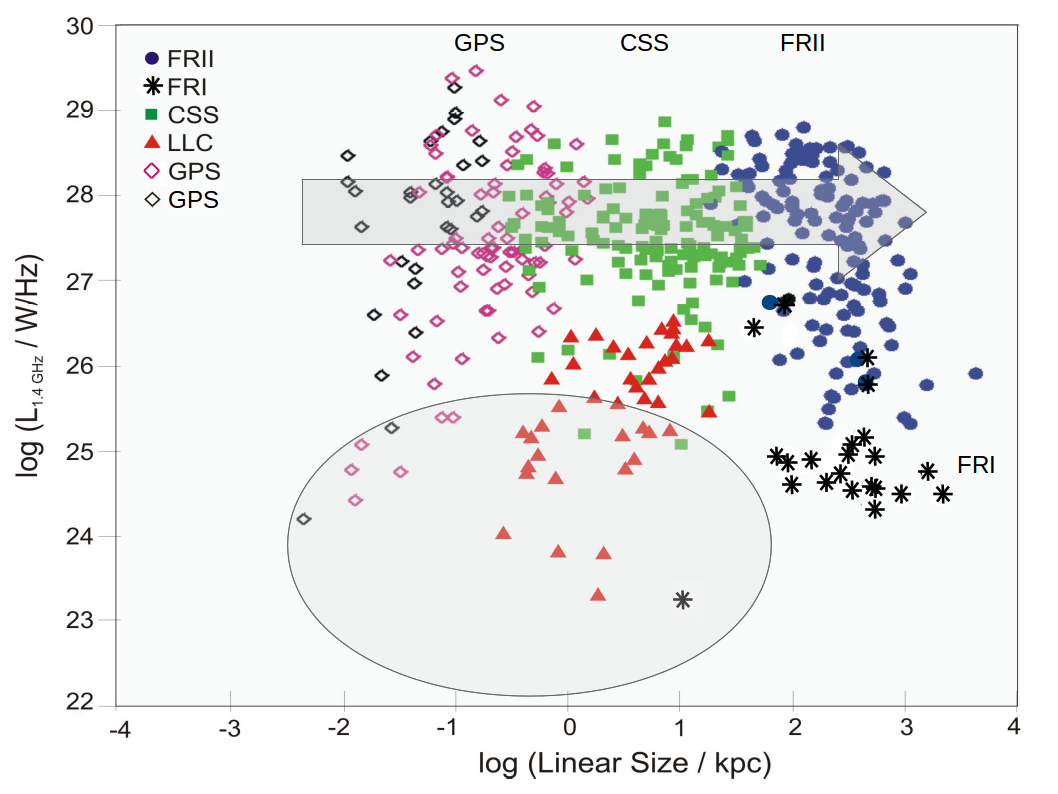}
\caption{\small \it  Radio power vs. linear size evolutionary scheme for radio-loud AGN (Kunert-Bajraszewska, M., Astronomische Nachrichten, 2016, Vol.337, Issue 1-2, p.27. Copyright Wiley-VCH Verlag GmbH \& Co. KGaA. Reproduced with permission). Squares represent CSS sources (Fanti et al. 2001, Marecki et al. 2003, Laing et al. 1983). Diamonds indicate GPS objects (Labiano et al. 2007). Stars indicate FRI and FRII objects (Laing et al. 1983).
Red triangles represent low luminosity compact (LLC) sources previously studied by Kunert-Bajraszewska $\&$ Labiano 2010, Kunert-Bajraszewska et al. 2010, 2014. The grey arrow indicates the main evolutionary trend of radio-loud AGNs. The oval indicates the boundaries of the area in which the new radio transient sources are located and where, as we expect, the whole population of short-lived objects can be found.
}
\end{figure}

In the general scenario of the evolution of powerful radio-loud AGNs (Fanti et al.1995), the younger and smaller Gigahertz-Peaked Spectrum (GPS) and Compact Steep Spectrum (CSS) sources become high luminosity large-scale FR\,II objects (Fanaroff \& Riley 1974; Fig.1). 
It seems reasonable to suspect that the compact AGNs with lower radio luminosity could be the progenitors of less luminous FR\,IIs and FR\,Is. However, the growing number of observations of low power radio sources and results of their analysis indicates that in their case the evolution can be more complicated or even halted at parsec scale. In order for the radio source to become a large-scale FR\,II or FR\,I object, the active phase needs to last longer than $10^4$ - $10^5$ years. The shorter active phase, which is probably caused by the low accretion rate, will result in poorly developed, sometimes disrupted, compact radio morphology. Some AGNs may undergo numerous short phases during their lifetime (Reynolds $\&$ Begelman 1997; Czerny et al. 2009, Kunert-Bajraszewska et al. 2010).
In this sense a 'young' GPS or CSS source means ongoing episode of the accretion disk outburst. 
This indicates the temporary existence of many weak CSS and GPS sources with compact or slightly resolved radio morphologies (core-jet) similar to that observed in radio-quiet objects (Giroletti \& Panessa 2009, Sadler et al. 2014). They are called short-lived radio objects on timescales $10^4$ - $10^5$ years. Interestingly, the enhancement and cessation of the accretion process is one of the postulated explanation for the origin of changing-look behaviour in the so-called 'changing look AGNs' (Elitzur et al. 2014, Schawinski et al. 2015). Such objects are rare and their first discoveries were made based on the X-ray variability. However, Koay et al. (2016) recently reported the changing-look behaviour in radio domain in one of such sources. This, in our opinion, links the discussion about the changing-look AGNs with the subject of short term radio activity.


To this day, only a handful of radio-loud short-lived AGN candidates have been found and studied (Kunert-Bajraszewska et al. 2010, 2014). However, the analysis of the radio, optical and X-ray proper-
ties of a sample of low-luminosity compact (LLC) radio sources carried out by our group over the last few years (Fig.1) suggests that a much larger population of such radio sources exist, and need to be explored.
The first unbiased study of these sources has been recently carried out with the Jansky VLA. The multi-year Caltech-NRAO Stripe 82 Survey (CNSS), revealed twelve transient sources on timescales of 5-20 years, largely associated with renewed AGN activity. The rates of such AGN possibly imply episodes of enhanced accretion and jet activity occurring once every 40,000 years in these galaxies. The whole CNSS consist of five epochs of observations over the entire $\sim 270\, \rm {deg}^{2}$ of Stripe 82 and they will be published soon.
The results from an initial pilot survey of a $\sim 50\, \rm {deg}^{2}$ of Stripe 82 has been already published by Mooley et al. 2016.




\begin{figure}[t]
\includegraphics[scale=0.4]{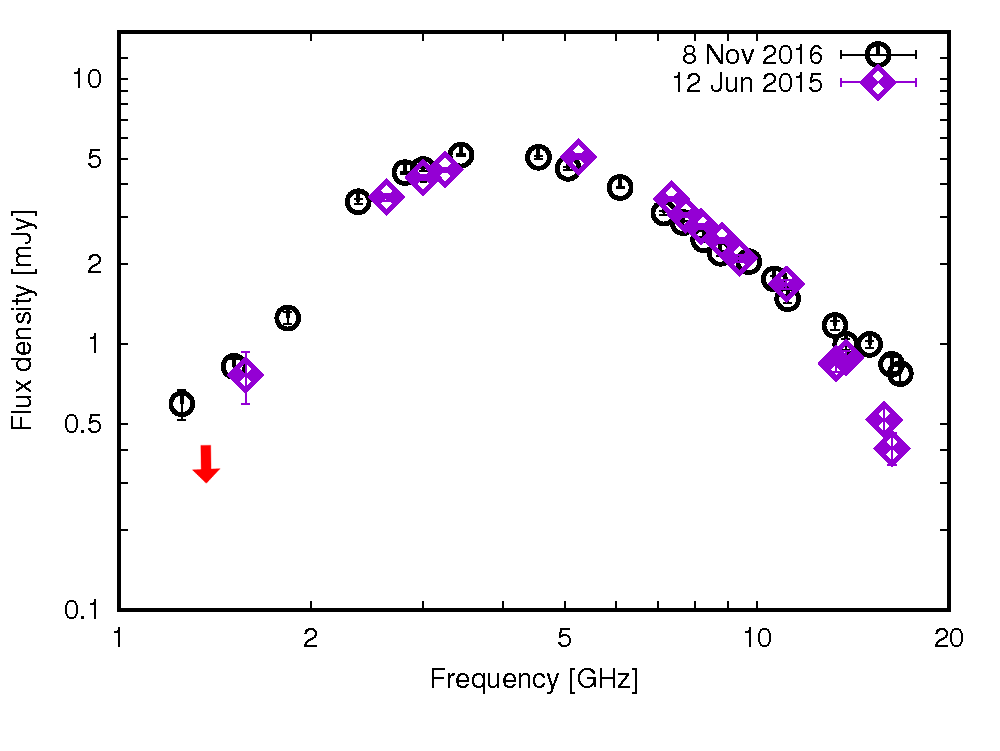}
\includegraphics[scale=0.38]{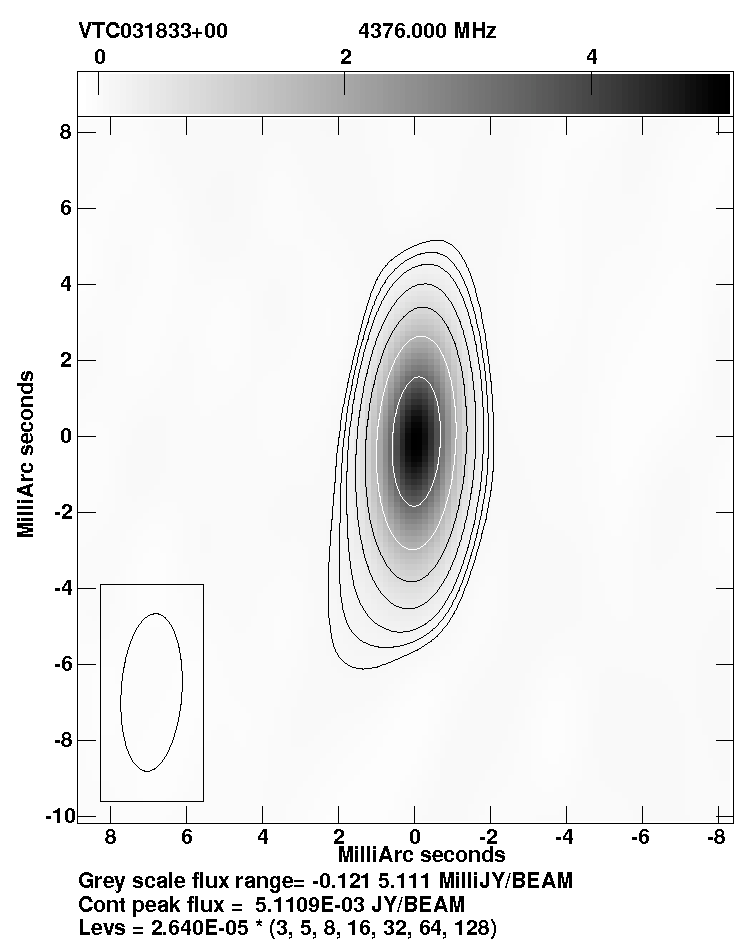}
\caption{\small \it The 1-20\,GHz VLA spectrum and 4.5\,GHz VLBA image of VTC031833+00. The sensitivity limit of the FIRST survey (1.4\,GHz, mean epoch 1999.2) is shown as a red arrow on the spectrum plot.}
\label{}
\end{figure}

\section{New radio transients}
The CNSS is a dedicated radio transient survey carried out with the Jansky VLA between December 2012 and May 2015. It was designed for systematically exploring the radio sky for slow transient phenomena on timescales between one day and several years. Observations of the 270 $deg^2$ of the SDSS Stripe 82 region were carried out over 5 epochs at 3 GHz and with a uniform rms noise of 80 $\mu Jy$ per epoch.
The CNSS survey has facilitated an unbiased study of short-lived sources for the first time. \\
\indent In this survey, 50 radio transients were discovered, among which are AGN, candidates of tidal disruption events and stellar explosions, flare stars, and active binary star systems (Mooley et al. 2016).
Although the majority of radio flaring phenomena in AGN are due to shocks propagating down the jets (e.g. Marscher \& Gear 1985), properties of some AGNs discovered as radio transients are distinct from this flaring population.\\
\indent This distinct population comprises AGNs that were not detected in radio in any previous survey of Stripe 82. They have been radio-quiet sources so far. After the outburst, the new radio sources are characterised by convex radio spectra peaking at a few GHz, which is typical for young AGNs - GPS sources. From our set of optical observations, we have determined the spectral types of these objects to be the second kind. In the type-II objects the central nucleus is obscured by a molecular-dusty torus, which means that the direction of view is close to the plane of the disk. This significantly reduces the
possibility of the Doppler enhancing of the jet flux fluctuations which can mimic the birth of new jet activity.

One such AGN, the type II quasar VTC233002-00 at redshift z=1.65 has been recently reported by Mooley et al. (2016). VTC233002-00 was detected at $\sim$5.5 mJy at 3 GHz in the CNSS, while the 3$\sigma$ upper limit at 1.4 GHz is 0.4 mJy (mean epoch 1999.2). The order-of-magnitude increase in flux density could be indicative of an enhanced accretion phenomenon leading to the production of a new jet. A comparison between the radio and optical flux densities indicates that this is a radio loud quasar. However, the value of the radio-loudness  parameter ($\rm logR$) calculated before and after the outburst changed from $\rm logR <1$ to $\rm logR = 2.1$, respectively. This indicates that VTC233002-00 has changed its status (look) from radio-quiet to radio-loud source.\footnote{We adopted radio-loudness definition from Kimball et al. 2011: $\rm logR = (M_{radio}-M_{i})/-2.5$,
where $\rm M_{radio}$ is a K-corrected radio absolute magnitude and $\rm M_{i}$ is a Galactic reddening corrected and K-corrected i-band absolute magnitude. Source is considered to be radio-loud if $\rm logR>1$.}

Finally, 12 such objects have been identified and we have undertaken a multi-frequency follow up campaign for those transient objects. New radio and X-ray observations and their analysis will be published soon. 
The latest preliminary results of our follow up VLA and VLBA observations of another of these objects are presented in Fig.2. VTC031833+00 has been discovered in 2015 in the CNSS survey and the Jansky VLA observations at 1-20 GHz carried by us in 2016 confirmed it is a GPS source peaking at 5GHz with a flux density of 5.1 mJy. The peak seems to slightly move towards lower frequency, when comparing the spectrum from June 2015 and the 2016 spectrum. That process could be due to propagating of the jet and its interaction with the circum-nuclear material in the host galaxy. The 4.5\,GHz VLBA image of VTC031833+00 shows slightly resolved radio structure, with a flux density of 5.1 mJy, consistent with published VLA results. This radio morphology is similar to those of the radio-quiet AGNs probably indicative of weak jets that are not able to develop large-scale structure (Ulvestad et al. 2005). 


\section{Summary}

Our sample represents a special class of radio-loud AGNs that harbor jets switched on within the past few decades.
These are, thus, the youngest radio sources that can provide information about the accretion state of supermassive black holes shortly after the onset of the jet formation. Furthermore, 
studies of interactions between the young radio source and interstellar medium can provide information about the energy that is deposited into the ISM by the expanding radio source. 
This is important to our understanding of the feedback process.
Taking this into account, we have undertaken a multi-frequency follow up campaign for the discovered transient objects. 
The detailed studies of their morphology and the jet evolution as well as the discussion about the postulated short term activity of AGNs will be published in forthcoming papers.

\section*{Conflict of Interest Statement}

The authors declare that the research was conducted in the absence of any commercial or financial relationships that could be construed as a potential conflict of interest.

\section*{Author Contributions}
All authors participated in acquisition, analysis and interpretation of data as well as the preparation of the manuscript.


\section*{Acknowledgments}
This is the contribution to the proceedings of the conference "Quasars at all cosmic epochs", held in Padua (Italy) in April 2017.

\section*{References}
\setlength{\parindent}{0cm}
Best, F. N., $\&$ Heckman, T. M.  2012, MNRAS, 421, 1569\\
Czerny, B., et al. 2009, ApJ, 698, 840\\
Elitzur, M., Ho, L. C., \& Trump, J. R. 2014, MNRAS, 438, 3340\\
Fanaroff, B.~L., \& Riley, J.~M. 1974, MNRAS 167, 31\\
Fanti, C., et al. 1995, A\&A 302, 317\\
Fanti  C., et al. 2001, A\&A, 369, 380\\
Giroletti, M., \& Panessa, F. 2009, ApJL, 706, L260\\
Kellerman, K., et al. 1989, AJ, 98, 1195\\
Kimball, A. E., et al. 2011, AJ, 141, 182\\
Koay, J.Y., et al. 2016, MNRAS, 460, 304\\ 
Kunert-Bajraszewska, M., et al. 2010, MNRAS, 408, 2261\\
Kunert-Bajraszewska, M., \& Labiano, A. 2010, MNRAS, 408, 2279\\
Kunert-Bajraszewska, M., et al. 2014, MNRAS, 437, 3063\\
Labiano A., et al. 2007, A$\&$A, 463, 97\\
Laing R.A., Riley J.M., Longair M.S. 1983, MNRAS, 204, 151\\
Marecki A., et al. 2003, PASA, 20, 42\\
Marscher, A. P., \& Gear, W. K. 1985, ApJ, 298, 114\\
Merloni, A., Heinz, S., \& di Matteo, T. 2003, MNRAS, 345, 1057\\
Mooley, K.P., et al. 2016, ApJ, 818, 105\\
Reynolds, C. S., \& Begelman, M. C. 1997, ApJ, 487, L135\\
Sadler, E.M., et al. 2014, MNRAS, 438, 796\\
Schawinski, K., et al. 2015, MNRAS, 451, 2517\\
Ulvestad, J.S., et al. 2005, ApJ, 621, 123


\end{document}